# Theoretical analysis and experimental verification on optical rotational Doppler effect

**Hailong Zhou,**[1,3] **Dongzhi Fu,**[2,3] **Jianji Dong,**[1,4] **Pei Zhang,**[2,5] **and Xinliang Zhang**[1]

[1]*Wuhan National Laboratory for Optoelectronics, School of Optical and Electronic Information, Huazhong University of Science and Technology, Wuhan 430074, China*
[2]*MOE Key Laboratory for Nonequilibrium Synthesis and Modulation of Condensed Matter, Department of Applied Physics, Xi'an Jiaotong University, Xi'an 710049, China*
[3]*These authors contributed equally to this work*
[4]*jjdong@hust.edu.cn*
[5]*zhangpei@mail.ustc.edu.cn*

**Abstract:** We present a theoretical model to sufficiently investigate the optical rotational Doppler effect based on modal expansion method. We find that the frequency shift content is only determined by the surface of spinning object and the reduced Doppler shift is linear to the difference of mode index between input and output orbital angular momentum (OAM) light, and linear to the rotating speed of spinning object as well. An experiment is carried out to verify the theoretical model. We explicitly suggest that the spatial spiral phase distribution of spinning object determines the frequency content. The theoretical model makes us better understand the physical processes of rotational Doppler effect, and thus has many related application fields, such as detection of rotating bodies, imaging of surface and measurement of OAM light.

## References and links

1. N. Seddon and T. Bearpark, "Observation of the inverse Doppler effect," Science **302**(5650), 1537–1540 (2003).
2. B. E. Truax, F. C. Demarest, and G. E. Sommargren, "Laser Doppler velocimeter for velocity and length measurements of moving surfaces," Appl. Opt. **23**(1), 67–73 (1984).
3. L. Marrucci, "Physics. Spinning the Doppler effect," Science **341**(6145), 464–465 (2013).
4. M. P. Lavery, F. C. Speirits, S. M. Barnett, and M. J. Padgett, "Detection of a spinning object using light's orbital angular momentum," Science **341**(6145), 537–540 (2013).
5. M. P. J. Lavery, S. M. Barnett, F. C. Speirits, and M. J. Padgett, "Observation of the rotational Doppler shift of a white-light, orbital-angular-momentum-carrying beam backscattered from a rotating body," Optica **1**(1), 1–4 (2014).
6. L. Allen, M. W. Beijersbergen, R. J. Spreeuw, and J. P. Woerdman, "Orbital angular momentum of light and the transformation of Laguerre-Gaussian laser modes," Phys. Rev. A **45**(11), 8185–8189 (1992).
7. S. Fürhapter, A. Jesacher, S. Bernet, and M. Ritsch-Marte, "Spiral interferometry," Opt. Lett. **30**(15), 1953–1955 (2005).
8. D. G. Grier, "A revolution in optical manipulation," Nature **424**(6950), 810–816 (2003).
9. J. E. Curtis and D. G. Grier, "Structure of optical vortices," Phys. Rev. Lett. **90**(13), 133901 (2003).
10. M. Chen, M. Mazilu, Y. Arita, E. M. Wright, and K. Dholakia, "Dynamics of microparticles trapped in a perfect vortex beam," Opt. Lett. **38**(22), 4919–4922 (2013).
11. A. Lehmuskero, Y. Li, P. Johansson, and M. Käll, "Plasmonic particles set into fast orbital motion by an optical vortex beam," Opt. Express **22**(4), 4349–4356 (2014).
12. G. Molina-Terriza, J. P. Torres, and L. Torner, "Twisted photons," Nat. Phys. **3**(5), 305–310 (2007).
13. A. Mair, A. Vaziri, G. Weihs, and A. Zeilinger, "Entanglement of the orbital angular momentum states of photons," Nature **412**(6844), 313–316 (2001).
14. N. Bozinovic, Y. Yue, Y. Ren, M. Tur, P. Kristensen, H. Huang, A. E. Willner, and S. Ramachandran, "Terabit-Scale Orbital Angular Momentum Mode Division Multiplexing in Fibers," Science **340**(6140), 1545–1548 (2013).
15. H. Zhou, J. Dong, L. Shi, D. Huang, and X. Zhang, "Hybrid coding method of multiple orbital angular momentum states based on the inherent orthogonality," Opt. Lett. **39**(4), 731–734 (2014).

16. J. Wang, J.-Y. Yang, I. M. Fazal, N. Ahmed, Y. Yan, H. Huang, Y. Ren, Y. Yue, S. Dolinar, M. Tur, and A. E. Willner, "Terabit free-space data transmission employing orbital angular momentum multiplexing," Nat. Photonics **6**(7), 488–496 (2012).
17. M. Padgett, "A new twist on the Doppler shift," Phys. Today **67**(2), 58–59 (2014).
18. J. Courtial, K. Dholakia, D. Robertson, L. Allen, and M. Padgett, "Measurement of the rotational frequency shift imparted to a rotating light beam possessing orbital angular momentum," Phys. Rev. Lett. **80**(15), 3217–3219 (1998).
19. C. Rosales-Guzmán, N. Hermosa, A. Belmonte, and J. P. Torres, "Experimental detection of transverse particle movement with structured light," Sci. Rep. **3**, 2815 (2013).
20. D. B. Phillips, M. P. Lee, F. C. Speirits, S. M. Barnett, S. H. Simpson, M. P. J. Lavery, M. J. Padgett, and G. M. Gibson, "Rotational Doppler velocimetry to probe the angular velocity of spinning microparticles," Phys. Rev. A **90**(1), 011801 (2014).
21. C. Rosales-Guzmán, N. Hermosa, A. Belmonte, and J. P. Torres, "Direction-sensitive transverse velocity measurement by phase-modulated structured light beams," Opt. Lett. **39**(18), 5415–5418 (2014).
22. G. Nienhuis, "Doppler effect induced by rotating lenses," Opt. Commun. **132**(1-2), 8–14 (1996).
23. I. V. Basistiy, V. V. Slyusar, M. S. Soskin, M. V. Vasnetsov, and A. Y. Bekshaev, "Manifestation of the rotational Doppler effect by use of an off-axis optical vortex beam," Opt. Lett. **28**(14), 1185–1187 (2003).
24. F. C. Speirits, M. P. J. Lavery, M. J. Padgett, and S. M. Barnett, "Optical angular momentum in a rotating frame," Opt. Lett. **39**(10), 2944–2946 (2014).
25. J. Leach, S. Keen, M. J. Padgett, C. Saunter, and G. D. Love, "Direct measurement of the skew angle of the Poynting vector in a helically phased beam," Opt. Express **14**(25), 11919–11924 (2006).
26. N. Zhang, J. A. Davis, I. Moreno, D. M. Cottrell, and X. C. Yuan, "Analysis of multilevel spiral phase plates using a Dammann vortex sensing grating," Opt. Express **18**(25), 25987–25992 (2010).
27. X. C. Yuan, J. Lin, J. Bu, and R. E. Burge, "Achromatic design for the generation of optical vortices based on radial spiral phase plates," Opt. Express **16**(18), 13599–13605 (2008).
28. Y. S. Rumala, "Sensitivity in frequency dependent angular rotation of optical vortices," Appl. Opt. **55**, 2024 (2016).
29. J. Lin, X. Yuan, S. H. Tao, and R. E. Burge, "Synthesis of multiple collinear helical modes generated by a phase-only element," J. Opt. Soc. Am. A **23**(5), 1214–1218 (2006).
30. T. W. Clark, R. F. Offer, S. Franke-Arnold, A. S. Arnold, and N. Radwell, "Comparison of beam generation techniques using a phase only spatial light modulator," Opt. Express **24**(6), 6249–6264 (2016).

## 1. Introduction

Linear Doppler effect is a well-known phenomenon by which the frequency of a wave is shifted according to the relative velocity of the source and the observer. This frequency shift scales with both the unshifted frequency and the linear velocity, and it is extensively used in Doppler velocimetry to detect the translational motion of surfaces and fluids [1, 2]. The Doppler velocimetry is a very mature technology and has been fully explored many years ago. In recent years, there has been increasing interest in another type of Doppler effect, namely rotational Doppler effect, in which a spinning object with an optically rough surface may induce a Doppler shift in light reflected parallel to the rotation axis, provided that the light carries orbital angular momentum (OAM) [3–5]. The OAM light comprises a transverse angular phase profile equal to $\exp(il\theta)$, where $\theta$ is the angular coordinate and $l$ is the azimuthal index, defining the topological charge (TC) of the OAM modes [6], namely the OAM mode index. These beams have an OAM of $l\hbar$ per photon ($\hbar$ is Planck's constant divided by $2\pi$) and consist of a ring of intensity with a null at the center. OAM light has been widely used in a variety of interesting applications, such as in optical microscopy [7], micromanipulation [8–11], quantum information [12, 13], optical communication [14–16]. And it has also been applied in probing the angular velocity of spinning microparticles or objects based on rotational Doppler effect [3–5, 17, 18].

In 2013, Padgett's group recognized that the well-known Doppler shift and Doppler velocimetry had an angular equivalent, and other works about measurement of transverse velocity based on rotational Doppler effect were also presented [3–5, 18–23]. In 2014, the physical mechanism of rotational Doppler effect was theoretically studied [24]. But this work was the case where the observer spun relative to the beam axis. A more general case is that a fixed input light illuminates a spinning object and the scattered light has a frequency shift. The mechanism of the frequency shift is still not very thorough. For example, it is still not clear how the surface of object influences the frequency shift content and what is the relationship between the incident light, scattered light and the surface.

In this paper, we theoretically investigate the optical rotational Doppler Effect using modal expansion method. We find that the frequency shift content is only determined by the surface of spinning object and the reduced Doppler shift of $(l-m)\Omega/2\pi$ is linear to the change of mode index, where $l$ is the mode index of the incident OAM light and $m$ is the one of the OAM light reflected or transmitted from a surface rotating at a fixed speed of $\Omega$. We design an experiment based on the beating effect to verify the theoretical model. We explicitly suggest that the design of spatial phase distribution of spinning object is crucial to generate the frequency shift content. The theoretical model makes us better understand the physical processes of rotational Doppler effect. It can provide theoretical guidance for many related applications, such as detection of rotating bodies, imaging of surface and measurement of OAM light.

## 2. Theoretical model

When an OAM light at frequency $f$ (wavelength $\lambda$) illuminates a spinning object with a rotating speed $\Omega$, as shown in Fig. 1, we choose the mode indexes of the incident and scattered light as $l$ and $m$ respectively. Here, the coordinate systems are always the same for the incident and scattered light. Without loss of generality, the following analysis is also applicable to the transmitted light. In a helically phased beam with $TC=l$, the skew angle between the Poynting vector and the beam axis is $l\lambda/2\pi r$, where $r$ is the radius from the beam axis [4, 5, 25]. So the incident angle and scattered angle are $\alpha = l\lambda/2\pi r$ and $\beta = m\lambda/2\pi r$ respectively. The reduced Doppler shift is

$$\Delta f = v(\sin\alpha - \sin\beta)f/c \approx (l-m)\Omega/2\pi, \qquad (1)$$

where $c$ is the speed of light in vacuum and $v = \Omega r$. The skew angle is very small because the radii of OAM modes are much larger than the light wavelength. Here, the approximation $\sin\alpha \approx \alpha, \sin\beta \approx \beta$ are used. So when an OAM light illuminates a spinning object with a rotating speed $\Omega$, the scattered light has a reduced frequency shift $(l-m)\Omega/2\pi$, which is related to the mode index difference of the incident and scattered light, and the rotating speed. Equation (1) is based on the fact that the surface is rough enough so that it can scatter the incident light in a mass of directions, i.e., the scattered light contains plentiful OAM modes. Although the rotational Doppler Effect is demonstrated from Eq. (1), it is still not clear how the surface of object influences the frequency shift content and the OAM modes.

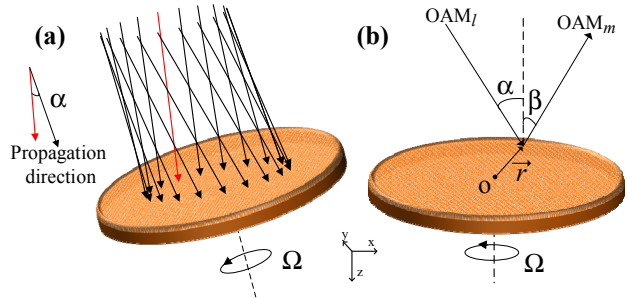

Fig. 1. (a) Schematic diagram and (b) analysis of rotational Doppler effect.

In the following, we will introduce a detail analysis to the relationship between the incident light, scattered light and the spinning surface by using modal expansion method. As we know, OAM mode conversion can be implemented by an ideal spiral phase plate (SPP), which can carry a helical phase $\exp(in\theta)$ on the illumination light [26–28]. As shown in Fig. 2(a), when an input OAM light with TC equal to $l$ (OAM$l$) illuminates on the SPP, the TC of the output OAM mode is changed to $l+n$. The similar principle can be applied to the rotational Doppler effect. The incident light and scattered light can be regarded as approximate parallel to the rotation axis because the skew angles are very small when the TCs of the OAM modes are not too large, or we can ignore the light with large skew angles because it is hard to collect. Meanwhile, the errors can be further reduced by collimating the incident light or increasing the beam size. We firstly assume that the reflectivity from the spinning object is homogeneous. It means that the spinning object can be regarded as a pure phase modulator and the modulated phase depends on the roughness of surface. The roughness of stationary surface can be written as $h(r,\theta)$ shown in Fig. 2(b), so the modulated phase is given by $\Phi(r,\theta)=4\pi h(r,\theta)/\lambda$. Similar to Ref [20], we rewrite the modulation function in Fourier expansion form as $\exp(i\Phi(r,\theta))=\sum A_n(r)\exp(in\theta)$, where $A_n(r)$ is the complex amplitude of $n$-order harmonic and $\sum |A_n(r)|^2=1$. Considering the spin, it is revised to

$$M(r,\theta)=\exp(i\Phi(r,\theta-\Omega t))=\sum A_n(r)\exp(in\theta)\exp(-in\Omega t), \qquad (2)$$

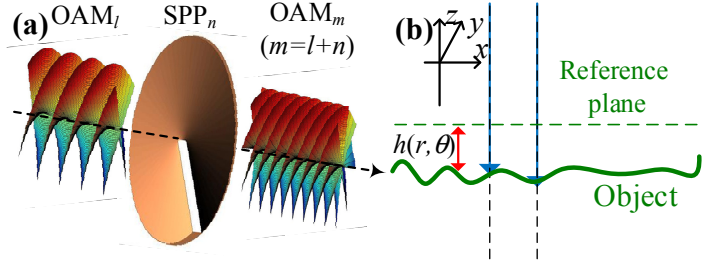

Fig. 2. (a) OAM mode conversion with a spiral phase plate. (b) Analysis on rough surface.

where $n$ are integers. So when an OAM mode at frequency $f$ expressed by $B(r)\exp(-i2\pi ft)\exp(il\theta)$ is normally incident on the spinning object, the scattered light can be expressed as

$$\sum B(r)A_n(r)\exp(-i2\pi ft)\exp(-in\Omega t)\exp(in\theta)\exp(il\theta). \qquad (3)$$

From Eq. (3), we can see that the scattered light is composed of many OAM modes with TCs equal to $m=n+l$, and every OAM mode has a different reduced frequency shift $(l-m)\Omega/2\pi$. These results conform to the previous analyses shown in Fig. 1 and Eq. (1). We can also see that the frequency shift content ($\Delta f=n\Omega/2\pi, n\in Z$) are only determined by the surface of spinning object and are not associated with the mode content of incident light. It means that as long as the surface of spinning object contains any helical phase component $\exp(in\theta)$ ($n\neq 0$), there will be frequency shifts whatever the mode content of incident light are. The incident light only determines the mode content of scattered light. So the non-twisted input OAM light (TC = 0) can also have a frequency shift. These theoretical predictions are consistent with the Marrucci's report [3].

More generally, when the reflectivity from the spinning object is not homogeneous, there are both amplitude modulation and phase modulation for the input light. The modulation function can be revised to $A(r,\theta)\exp(i\Phi(r,\theta))$, where $A(r,\theta)$ is corresponding to the amplitude modulation. It can also be rewritten in Fourier expansion form similar to Eq. (2).

In order to verify the theoretical model, we design our experiment based on the beating effect. When only two OAM modes, expressed by $\sum_{s=1,2} B_s(r)\exp(il_s\theta)\exp(-i2\pi ft)$, shine the spinning object, the scattered light is deduced as

$$E_o = \sum_{m\in Z}\begin{Bmatrix} B_1(r)\exp(-i2\pi ft)A_{m-l_1}(r)\exp(im\theta)\exp[-i(m-l_1)\Omega t]+ \\ B_2(r)\exp(-i2\pi ft)A_{m-l_2}(r)\exp(im\theta)\exp[-i(m-l_2)\Omega t]+ \end{Bmatrix}. \quad (4)$$

After spatial filtering, only the fundamental mode (TC = 0) is selected and then is collected by a photodetector (PD). The collected intensity dependent on the time can be derived as

$$I(t) = \iint \sum_{s=1,2,} \left|B_s(r)A_{-l_s}(r)\right|^2 rdrd\theta + \\ 2\iint rdrd\theta \left|B_1(r)A_{-l_1}(r)B_2(r)A_{-l_2}(r)\right|\cos[(l_1-l_2)\Omega t + \phi(r)], \quad (5)$$

where $\phi(r)=\mathrm{angle}\left[B_1(r)A_{-l_1}(r)B_2^*(r)A_{-l_2}^*(r)\right]$. Ignoring the difference along the radial, Eq. (5) can be simplified to

$$I(t) = \left|A_{-l_1}\right|^2 I_1 + \left|A_{-l_2}\right|^2 I_2 + 2\left|A_{-l_1}A_{-l_2}\right|\sqrt{I_1 I_2}\cos[(l_1-l_2)\Omega t + \phi], \quad (6)$$

where $I_1$ and $I_2$ are the power of the two input OAM modes. From Eq. (6), we can see that the amplitude of beating signal at $(l_1-l_2)\Omega/2\pi$ is linear with the one of $-l_1$ and $-l_2$-order harmonics of spinning object. So we can fix one of the OAM mode as reference mode and scan another mode to measure the harmonic distribution of spinning object.

## 3. Experiment results

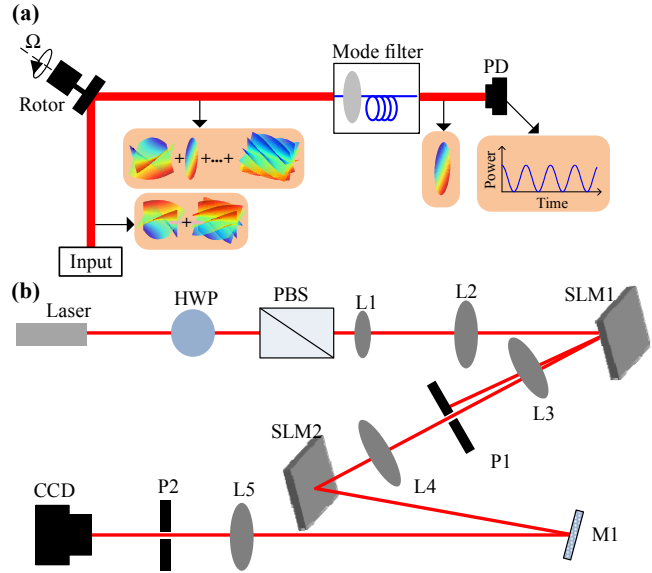

Fig. 3. (a) Schematic and (b) experimental setup to measure the beating signal.

Figure 3(a) shows the schematic of experimental design. Two OAM modes are incident on the spinning object. Modulated by the rough surface, the light scattered from the spinning object contains many OAM modes. After a mode filter, the fundamental mode is selected and then collected by a PD. Owing to the rotational Doppler Effect, a beating signal is measured. In our experiment, we employ a spatial light modulator (SLM1) to generate the two OAM modes and another one (SLM2) to emulate the spinning object by rotating the pattern of SLM2. Figure 3(b) shows the experimental setup. The light at 633 nm emitted from the He–Ne laser is expanded with two lenses (L1 and L2), and then illuminates SLM1. The half wave plate (HWP) and polarization beam splitter (PBS) are used to tune the input power and polarization. SLM1 is used to generate the two OAM modes. The OAM modes are filtered out from the first-order diffracted beam with a pinhole (P1) and then illuminates the spinning object (SLM2). The pattern of SLM2 is spined by software to scan the azimuth. Afterwards, the fundamental mode is selected by another pinhole (P2) and then a charge-coupled device (CCD) is used to collect the intensity. The lenses (L3, L4, and L5) and mirror (M1) are used to tune the beam size and optical route. The SLM can realize an arbitrary modulation function shown in Eq. (2) by different algorithms [29, 30]. Here, the spinning object is replaced with an SLM by rotating the pattern of SLM.

We firstly add a random pattern in SLM2 and input two OAM modes to verify the beating effect. Figure 4(a) shows the simulated interference patterns and experimental results of OAM0 mixed with OAM-6, OAM0 mixed with OAM-3 and OAM0 mixed with OAM3 respectively. We can see that they match well each other. The measured intensity dependent on rotation angle in one period is presented in Fig. 4(b), when the two input OAM modes are OAM0 and OAM4. Then we calculate power spectrum, as shown in Fig. 4(c). Obviously, a beating signal at 4 times rotational speed is measured. Using this effect, we can measure the harmonic distribution of the spinning object to verify our theoretical model.

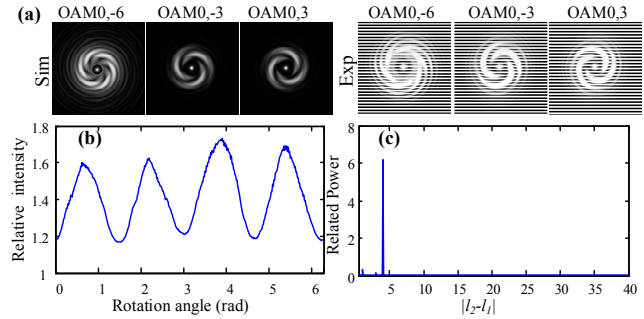

Fig. 4. (a) Simulated interference patterns and experimental results of different states. (b) The measured intensity depend on rotation angle in one period and (c) the calculated power spectrum, when the two input OAM modes are OAM0 and OAM4.

We design the pattern of SLM2 to have a binary harmonic distribution shown in Fig. 5(a). The total power is normalized to unit. The blue bars denote the ideal weight of each component and the red bars denote the theoretical results by using iterative algorithm [29]. We can see that they are basically accordant. Then we fix the power of the two input OAM modes. We select the OAM0 (TC = 0) as the reference mode and scan another mode from OAM4 (TC = 4) to OAM-12 (TC = −12). According to our theoretical model, the measured power of every beating signal is proportional to the power weight of corresponding harmonics of the spinning object. Figure 5(b) shows the measured harmonic distributions compared to the theoretical results. The blue bars denote the theoretical results and the red bars denote the experimental results. We can see that they agree well each other. It proves that the rotational Doppler Effect is decided by the spinning object. That is to see if the spinning object does not have the $n$-order harmonic component, there will be not the reduced frequency shift $\Delta f = n\Omega / 2\pi$ and vice versa. In fact, the power of output light, who has a reduced frequency shift of $n\Omega / 2\pi$, is proportional to the weight of $n$-order harmonic component. We design another pattern of SLM2 to have a normal harmonic distribution shown in Fig. 6(a). The strong −7-order harmonic component is to ensure good performance when choosing OAM7 as the reference mode. Figure 6(b) shows the measured harmonic distributions compared to the theoretical results by scanning another mode from OAM6 to OAM-6. We can see that the power of beating signal is precisely proportional to the weight of harmonic component. It provide strong evidence that the power of beam who has a reduced frequency shift of $n\Omega/2\pi$ is proportional to the weight of $n$-order harmonic component, which is consistent with our theoretical model.

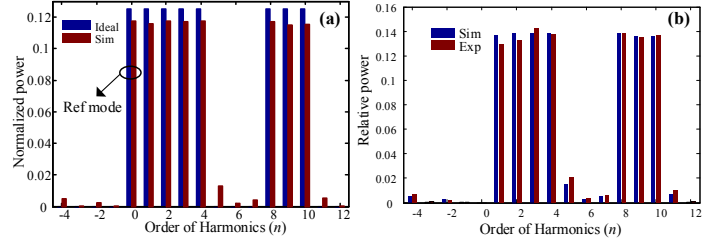

Fig. 5. (a) Binary harmonic distributions of the spinning object. The blue bars denote the ideal distributions and the red bars denote the theoretical results by using iterative algorithm. (b) The measured harmonic distributions compared to the theoretical results. The blue bars denote the theoretical results and the red bars denote the experimental results.

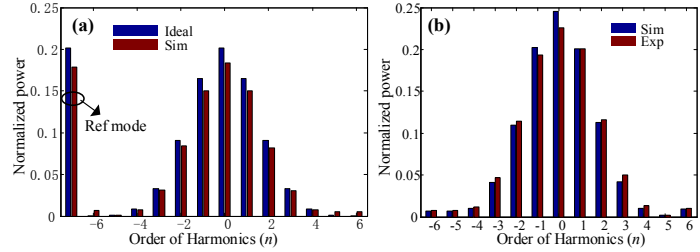

Fig. 6. (a) Normal harmonic distributions of the spinning object. (b) The measured harmonic distributions compared to the theoretical results.

## 4. Conclusion

In conclusion, we theoretically investigate the optical rotational Doppler Effect by using modal expansion method. We find that the frequency shift content is only determined by the surface of spinning object and the reduced Doppler shift is linear to the change of OAM mode index. Besides, we carry out a beating frequency experiment to verify the theoretical model. The experiment suggests that the power of beating signal is proportional to the weight of $n$-order harmonic component of spinning object, which is consistent with our theoretical model. Our theoretical model makes us better understand the physical processes of rotational Doppler Effect. We can optimize the parameters based on this model to get better performance for many applications of rotational Doppler Effect. It can provide theoretical guidance for many related applications, such as detection of rotating bodies, imaging of surface and measurement of OAM.

**Acknowledgment**

This work was partially supported by the Program for New Century Excellent Talents in Ministry of Education of China (Grant No. NCET-11-0168), a Foundation for the Author of National Excellent Doctoral Dissertation of China (Grant No. 201139), the National Natural Science Foundation of China (NSFC) (Grant No. 11174096, 11374008 and 61475052), Foundation for Innovative Research Groups of the Natural Science Foundation of Hubei Province (Grant No. 2014CFA004).